\documentstyle{mn}
\input{psfig}
\newcommand{\etal}{{et al.~}}

\newcommand{\kmsmpc}{\>{\rm km}\,{\rm s}^{-1}\,{\rm Mpc}^{-1}}

\newcommand{\Mpc}{\>{\rm Mpc}}

\newcommand{\Msun}{\>{\rm M_{\odot}}}

\newcommand{\beq}{\begin{equation}}
\newcommand{\eeq}{\end{equation}}
\newcommand{\Rvir}{r_{\rm vir}}

\def\mpchi{\,h^{-1}{\rm {Mpc}}}

\def\gtsima{$\; \buildrel > \over \sim \;$}
\def\ltsima{$\; \buildrel < \over \sim \;$}
\def\prosima{$\; \buildrel \propto \over \sim \;$}
\def\gsim{\lower.7ex\hbox{\gtsima}}
\def\lsim{\lower.7ex\hbox{\ltsima}}
\def\simgt{\lower.7ex\hbox{\gtsima}}
\def\simlt{\lower.7ex\hbox{\ltsima}}
\def\simpr{\lower.7ex\hbox{\prosima}}
\def\la{\lsim}
\def\ga{\gsim}
\def\lta{\la}
\def\gta{\ga}

\newcommand{\apj}{ApJ}
\newcommand{\apjs}{ApJS}

\newcommand{\mnras}{MNRAS}
\newcommand{\aap}{A\&A}

\newdimen\hssize
\newdimen\hdsize
\begin{document}


\title[The Alignment between Satellites and Central Galaxies]
      {The Alignment between Satellites and Central Galaxies: 
       Theory vs. Observations}

\author[Kang et al.]
       {\parbox[t]{\textwidth}{
        X. Kang$^{1,2}$\thanks{E-mail:kang@mpia.de}, 
        Frank C. van den Bosch$^{1}$, Xiaohu Yang$^{3}$, Shude Mao$^{4}$, 
        H.J. Mo$^{5}$,\\ Cheng Li$^{3,6}$, Y.P. Jing$^{3}$}\\
        \vspace*{3pt} \\
        $^1$Max-Planck-Institute for Astronomy, K\"onigstuhl 17, D-69117
            Heidelberg, Germany\\
        $^2$Astrophysics, University of Oxford, Denys Wilkinson Building, 
            Keble Road, Oxford OX1 3RH, UK\\
        $^3$Shanghai Astronomical Observatory; the Partner Group of MPA,
            Nandan Road 80,  Shanghai 200030, China\\
        $^4$University of Manchester, Jodrell Bank Observatory,
            Macclesfield, Cheshire SK11 9DL, UK\\
        $^5$Department of Astronomy, University of Massachusetts,
            Amherst MA 01003-9305, USA\\
        $^6$Max-Planck-Institute for Astrophysics, Karl Schwarzschild
            Str. 1, D-85741 Garching, Germany}


\date{}

\pagerange{\pageref{firstpage}--\pageref{lastpage}}
\pubyear{2000}

\maketitle

\label{firstpage}


\begin{abstract}
  Recent  studies  have  shown  that  the  distribution  of  satellite
  galaxies  is preferentially  aligned with  the major  axis  of their
  central galaxy.   The strength of  this alignment has been  found to
  depend  strongly  on  the  colours  of  the  satellite  and  central
  galaxies,  and only  weakly on  the mass  of the  halo in  which the
  galaxies  reside.  In this  paper we  study whether  these alignment
  signals, and their dependence on  galaxy and halo properties, can be
  reproduced  in  a  hierarchical   structure  formation  model  of  a
  $\Lambda$CDM concordance  cosmology.  To that extent we  use a large
  $N$-body  simulation which  we  populate with  galaxies following  a
  semi-analytical  model for galaxy  formation.  We  find that  if the
  orientation of the central galaxy  is perfectly aligned with that of
  its dark matter halo, then the predicted central-satellite alignment
  signal is much stronger than  observed.  If, however, the minor axis
  of a central  galaxy is perfectly aligned with  the angular momentum
  vector  of its  dark matter  halo, we  can accurately  reproduce the
  observed  alignment strength  as function  of halo  mass  and galaxy
  color.   Although  this suggests  that  the  orientation of  central
  galaxies is  governed by the  angular momentum of their  dark matter
  haloes, we emphasize that any other scenario in which the minor axes
  of central galaxy  and halo are misaligned by  $\sim 40^{\circ}$ (on
  average) will  match the data  equally well.  Finally, we  show that
  dependence of  the alignment  strength on the  color of  the central
  galaxy is  most likely an artefact  due to interlopers  in the group
  catalogue.  The  dependence on the color of  the satellite galaxies,
  on the other hand, is real  and owes to the fact that red satellites
  are associated with  subhaloes that were more massive  at their time
  of accretion.
\end{abstract}




\section{Introduction}
\label{sec:intro}

Satellite galaxies are a useful tracer of the dark matter distribution
on  the  scale of  individual  dark  matter  haloes.  Since  they  are
typically  distributed over  the  entire dark  matter  halo, they  are
ideally suited as  a tracer population of the  potential well in which
they orbit.   In particular,  their kinematics can  be used  to obtain
accurate dynamical  masses of  their host haloes  (e.g., McKay et  al. 
2002;  Brainerd \& Specian  2003; Prada  et al.   2003; van  den Bosch
\etal  2004; Conroy  \etal 2007).   In addition,  one can  also obtain
useful  constraints  from  the   radial  and  angular  number  density
distributions   of  satellite   galaxies.   For   example,   a  strong
photoionizing  background  may  strongly  suppress star  formation  in
low-mass (sub)haloes  (e.g Quinn,  Katz, \& Efstathiou  1996; Bullock,
Kravtsov  \& Weinberg 2000;  Benson \etal  2002).  Hence  the observed
number of  satellite galaxies  can be used  to put constraints  on the
efficiency of  this process.  Furthermore,  numerical simulations have
shown that dark  matter haloes are in general  triaxial (e.g., Jing \&
Suto  2002)  and  that  the  massive  progenitors  are  preferentially
accreted along the direction of the large-scale filaments (Knebe \etal
2004;  Benson 2005;  Wang \etal  2005; Zentner  \etal  2005; Libeskind
\etal  2005).    Consequently,  the  {\it   angular}  distribution  of
satellite  galaxies  contains  information  regarding  the  shape  and
orientation of dark matter haloes. This is the topic of this paper.

The  alignment of central galaxies   with satellite galaxies was first
studied  nearly four  decades ago  by Holmberg  (1969), who found that
satellites are preferentially located along the minor axes of isolated
disc galaxies.  There have  since been  many studies with  conflicting
results (e.g.,  Hawley \&  Peebles  1975; Sharp,  Lin  \& White  1979;
Zaritsky \etal 1997), mainly because the  samples used were relatively
small.  With the advent of large galaxy  surveys, such as the 2-degree
Field Galaxy Redshift Survey (2dFGRS) and the Sloan Digital Sky Survey
(SDSS),  much  larger  samples  can   now   be constructed   to  study
central-satellite  alignments.   Recent  results   from  such  samples
demonstrated    that  satellite galaxies   are  in fact preferentially
aligned with  the {\it major} axis of  the central  galaxies (Brainerd
2005; Yang  \etal  2006;  Azzaro \etal   2006)\footnote{An independent
  study by Sales   \& Lambas (2004)  based  on the 2dFGRS, claimed  to
  detect a minor axis alignment.  However, due to  an error with their
  definition of  the orientations  angles,  they actually   detected a
  major axis alignment   (see   discussion  in Yang   \etal   2006).}.
Brainerd (2005) studied  a sample of  isolated SDSS galaxies and found
that the satellites around these  galaxies are preferentially oriented
along their major axes.  Yang \etal (2006;  hereafter Y06) studied the
alignment between  the  central   galaxies and  the    distribution of
satellite galaxies in  a large sample  of galaxy  groups selected from
the SDSS, and confirmed  the preference for  major axis  alignment. In
addition, Y06 found that the alignment signal depends  on the color of
the central and  satellite   galaxies:  it is strongest  between   red
centrals and red satellites,  and almost absent between blue  centrals
and  blue satellites.  These results,  in turn, have been confirmed by
Azzaro \etal  (2006),    who   studied the  alignments   in   isolated
host-satellite systems in  the  SDSS.   Finally,  Y06 found that   the
strength of  the alignment signal   increases  weakly with  increasing
group (halo) mass.
 
The goal of this paper  is to examine whether these observed alignment
signals  can be  reproduced  in the  hierarchical structure  formation
model. To  that extent  we use a  large $N$-body simulation  which has
been  populated with  galaxies following  a semi-analytical  model for
galaxy formation.  Our study is similar in spirit to that of Agustsson
\&   Brainerd  (2006,  hereafter   AB06).   However,   our  simulation
resolution (see \S\ref{sec:method})  is significantly higher than that
used  by AB06,  allowing us  to include  satellite galaxies  with much
lower masses.  In addition, we  use very different  (less restrictive)
host-satellite  selection criteria,  and we  also investigate  how the
alignment strength  depends on galaxy  color and halo  mass, something
that was not addressed by AB06. Finally, when comparing our simulation
results with observations, we use realistic mock catalogs to take into
account various selection effects (see \S\ref{sec:mis}).

The outline of  the  paper is as  follows.   In  \S\ref{sec:method} we
present our numerical simulations,  and describe how the orientations
of central  galaxies are  defined and how   the alignment signals  are
analyzed.  In \S\ref{sec:align}   we explore the alignments  under the
hypothesis that the central galaxy is oriented along the major axis of
the dark matter halo in  projection.  In \S\ref{sec:mis} we  construct
more realistic   mock catalogs, which   we  use  to examine   how  the
alignment signal changes due to selection effects. We also investigate
two more realistic models  for the orientation  of the central galaxy;
one  based on the  inertia moment of its halo,  the other based on the
halo angular momentum.  We summarize our results In \S\ref{sec:concl}.

\section{Methodology}
\label{sec:method}

The numerical simulation and  semi-analytical model used in this paper
are described in detail in Kang \etal (2005a; hereafter K05) and Kang,
Jing \&  Silk (2006).  Below we  give a brief description  of the main
ingredients of the model, and we  refer the reader to these papers for
more details.

\subsection{$N$-body simulations and the semi-analytical model}  
\label{sec:Nbody}

The numerical  simulation used  here has been  carried out by  Jing \&
Suto (2002)  using a vectorized-parallel P$^3$M code.   It follows the
evolution  of  $512^{3}$  particles  in  a cosmological  box  of  $100
\mpchi$,  assuming a  flat $\Lambda$CDM  `concordance'  cosmology with
$\Omega_m =0.3$, $\sigma_{8}=0.9$, and $h=(H_0/100\kmsmpc)=0.7$.  Each
particle has a mass of  $6.2 \times 10^{8} h^{-1} \Msun$.  Dark matter
haloes  are identified  using the  friends-of-friends  (FOF) algorithm
with a linking length equal to 0.2 times the mean particle separation.
For each halo  thus identified we compute the  virial radius, $\Rvir$,
defined  as the  {\it spherical}  radius inside  of which  the average
density is $101$ times the critical density of the Universe (cf. Bryan
\& Norman 1998). The virial mass  is simply defined as the mass of all
particles that have halocentric radii  $r \leq \Rvir$, and is about 15
percent  smaller than  the FOF  mass. Note  that $\Rvir$  is  about 30
percent larger than  the often used radius $r_{200}$,  inside of which
the  average  density is  $200$  times  the  critical density  of  the
universe.   In   our  alignment  analysis,   we  will  use   both  the
(non-spherical) friends-of-friends haloes,  and the (spherical) virial
haloes, hereafter FOF and VIR haloes, respectively.

Dark  matter subhaloes within  each FOF  (parent) halo  are identified
using the SUBFIND  routine described in Springel \etal  (2001). In the
present study,  we use  all haloes and  subhaloes with masses  down to
$6.2   \times  10^{9}h^{-1}M_{\odot}$   (10   particles).   Using   60
simulation  outputs  between  $z=15$  and  $z=0$,  equally  spaced  in
$\log(1+z)$, K05 constructed the  merger history for each (sub)halo in
the simulation box, which are  then used in the semi-analytical model. 
In  what  follows, whenever  we  refer  to a  {\it  halo},  we mean  a
virialized object which is not  a sub-structure of a larger virialized
object, while {\it subhaloes} are virialized objects that orbit within
a halo.

In the semi-analytical model it is  assumed that the baryonic gas in a
halo is heated to the  virial temperature of the halo by gravitational
shocks.  Subsequently, this hot  gas cools, radiating away its binding
energy, and settles down into the center of the halo to form a central
galaxy (White  \& Rees 1978). The  star formation rate in  a galaxy is
assumed  to  be proportional  to  the total  amount  of  cold gas  and
inversely  proportional  to the  dynamical  time  of  the system.   An
initial mass function  is assumed to estimate the  supernova rate, and
is combined  with a population  synthesis model and a  dust extinction
model to  calculate the luminosities  in different photometric  bands. 
In addition, our semi-analytical model also accounts for feedback from
supernova explosions  and active galactic nuclei (see  Kang \etal 2006
for details).
 
In this  model, each halo  contains a galaxy  at its center,  which we
call the {\it  central galaxy}, and which is assumed  to have the same
position  and  velocity  as the  most  bound  particle  of its  halo.  
Subhaloes also  host galaxies  at their center,  to which we  refer as
{\it halo galaxies}.   They are assigned the position  and velocity of
the most  bound particle  of their subhalo.   Note that  halo galaxies
were  central galaxies  before their  host haloes  fell into  a larger
halo.  When this  happens, the  hot  gas associated  with the  smaller
progenitor is assumed  to be stripped from its  halo, and becomes part
of the  hot gas reservoir of  the new parent  halo. Consequently, halo
galaxies are  no longer fed  by a cooling  flow of new gas,  and their
star formation terminates as soon  as their cold gas reservoir is used
up.

Subhaloes are subject to tidal mass loss while they orbit their parent
halo.  Consequently, subhaloes may be tidally disrupted, or may become
too small to be identified by the SUBFIND routine. If this happens its
halo galaxy is  attached to the most bound particle  of its subhalo at
the time  just before  it disappeared from  the (sub)halo  catalog. In
what follows we refer to  these galaxies as {\it orphan galaxies}. The
dark matter mass of the {\it orphan galaxies} are defined as the total
dark matter  mass of the subhalos  at the time they  are identified as
the halo galaxies.  The stellar  mass of the {\it orphan galaxies} are
determined according  to the star  formation evolution model  given in
detail  by K05.   The motion  of  an orphan  galaxy is  assumed to  be
governed by  dynamical friction, and  it is therefore merged  with the
central galaxy in the halo  after a dynamical friction time scale. The
combined set  of halo and orphan  galaxies which are found  in VIR (or
FOF) haloes are now referred to as {\it satellite galaxies}.

\subsection{The orientation of central galaxies}
\label{sec:orient}

We now describe  how we use the semi-analytical  model described above
to compute the alignment between the orientation of the central galaxy
and the distribution  of its satellite galaxies.  In  order to measure
the required alignments,  we first need to specify  the orientation of
the central  galaxies.  Unfortunately, the  semi-analytical model does
not make any predictions regarding either the shape or the orientation
of the galaxies.  Rather, these have  to be put in by hand.  We follow
AB06 and assume  that the orientation of central  galaxies is governed
by properties of their dark  matter haloes.  In this paper we consider
three scenarios: (i) the orientation  of the central galaxy is aligned
with  the {\it  projected} major  axis of  its dark  matter  halo (see
\S\ref{sec:align}),  (ii) the  minor  axis of  the  central galaxy  is
aligned with the {\it true}, 3-dimensional (3D) minor axis of the dark
matter halo  (see \S\ref{sec:mis}),  and (iii) the  minor axis  of the
central galaxy is aligned with the angular momentum vector of the dark
matter halo  (see \S\ref{sec:mis}).  Note that scenarios  (i) and (ii)
are  equivalent for  spherical  and axisymmetric  (oblate or  prolate)
haloes,  but  not for  the  more  general  triaxial haloes.  In  fact,
scenario (i) is  unphysical for triaxial haloes, as  it depends on the
projection axis used (i.e., on  the orientation of the line connecting
the galaxy and the observer). Clearly, the orientation of a galaxy can
not depend on this. The reason for nevertheless adopting this model is
twofold.  First of  all, it maximizes the alignment  signal, making it
useful to explore  various trends.  Secondly, the same  model was also
used by  AB06, thus  allowing for a  meaningful comparison  with their
results.  As  discussed above, we will consider  two halo definitions,
FOF and VIR. In each case,  we define the orientations of the halo and
the central galaxy using only those particles that are part of the FOF
or VIR halo in question.

In order to obtain the orientation of a dark  matter halo we determine
the principal axes of the  inertia tensor of  the distribution of dark
matter particles. We define the inertia tensor as
\begin{equation}
\label{eq:unweighted}
I_{ij} \equiv \sum_{n} x_{i,n} x_{j,n}
\end{equation}
where $x_{i,n}$ is the position of the $n^{\rm th}$ particle. In order
to reduce  the impact of  (massive) subhaloes on the  determination of
$I_{ij}$ the summation is over those  particles in the FOF or VIR halo
that are  not part of  a subhalo.  In  the case of the  projected mass
distribution,  $i,j =1,2$,  while  in 3D  configuration  space we  use
$i,j=1,2,3$. The  eigenvectors of  $I_{ij}$ define the  orientation of
the  halo,  while the  corresponding  eigenvalues  determine the  halo
shape.

Finally, we compute the angular momentum vector of each halo using
\begin{equation}
\label{eq:unweighted.J}
{\bf J} = \sum_{n} {\bf r}_{n} \times ({\bf v}_{n} - \bar{\bf v})
\end{equation}
Here ${\bf r}$ and ${\bf v}$  are the position and velocity vectors of
the dark matter particles, $\bar{\bf  v}$ is the mean bulk velocity of
the halo, and the summation is again over all dark matter particles of
the FOF or VIR halo in question  that are not part of a subhalo. Tests
have  shown, however,  that none  of our  alignment  results presented
below are  sensitive to whether  we remove these subhalo  particles or
not.

\subsection{Quantifying Alignment}
\label{sec:quant}

Having defined the orientation of  central galaxies, we now proceed to
quantify  the angular  distribution of  their satellite  galaxies. The
satellites  are defined  as all  the  halo galaxies  and {\it  orphan}
galaxies found  within $\Rvir$  (or the FOF  halo).  We start  by only
selecting those galaxies in haloes  with $M \geq 10^{12} h^{-1} \Msun$
and with  $M_{b_J} \leq -16$.  Here $M$  can be either the  FOF or the
virial mass,  depending on  what halo definition  we adopt.   The mass
limit  is  imposed to  ensure  that we  have  a  sufficient number  of
particles ($N > 1600$) to reliably measure the halo shape, orientation
and angular momentum.   In addition, the results of  Y06, which we use
for  comparison, have  also been  restricted  to haloes  with $M  \geq
10^{12}  h^{-1} \Msun$.   The  absolute magnitude  limit reflects  the
minimum  luminosity for  which the  SAM can  reproduce  the luminosity
function: the  simulation box  used does not  resolve the  halo masses
that  typically host  central galaxies  with $M_{b_J}  >  -16$.  These
selection   criteria  yield  $3746$   central  galaxies   and  $29941$
satellites, of  which 58 percent are {\it  orphan galaxies}.  Finally,
we  split the  galaxy  population  in two  subsamples  based on  their
photometric   color.    Following  Y06,   we   define  galaxies   with
$^{0.1}(g-r)\ge 0.83$ as {\it red} galaxies and the rest as {\it blue}
galaxies.  Here  $^{0.1}(g-r)$ is  the color in  the SDSS $g$  and $r$
bands, $K$-corrected to $z=0.1$.
\begin{figure}
\centerline{\psfig{figure=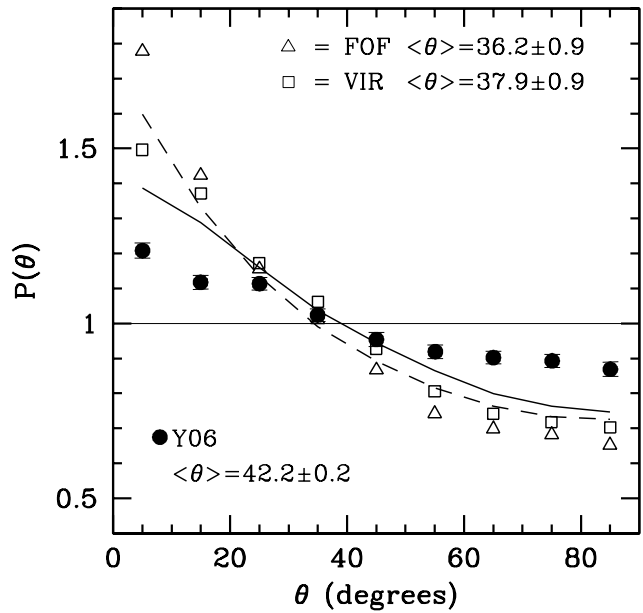,width=\hssize}}
\caption{The  normalized probability distribution, $P(\theta)$, of the
  angle $\theta$ between the major  axis of the central galaxy and the
  direction of each satellite as measured from the central galaxy. The
  open squares and  open triangles show the results  obtained from our
  SAM using  VIR and FOF  haloes, respectively.  The solid  and dashed
  lines show $P(\theta)$ for the  dark matter particles in VIR and FOF
  haloes.   Note that  the  satellite  galaxies in  the  SAM reveal  a
  stronger alignment than the  dark matter.  For comparison, the solid
  dots with  errorbars are the observational results  obtained by Yang
  \etal  (2006) from  a SDSS  galaxy group  catalogue.  Note  that the
  alignment signal in the SAM is much stronger than observed.}
\label{fig:all}
\end{figure}

In  our model,  the total  radial distribution  of the  satellites are
found to be similar with  the dark matter particles, but halo galaxies
are found to be more less  concentrated in the center of the halo, and
such  distributions  are  also  found  by other  works  (e.g.  Gao  et
al. 2004, Kang  et al. 2005, AB06).  These  For each central-satellite
pair  we compute  the  angle $\theta$,  defined  as the  angle on  the
projection plane between the position  of the satellite galaxy and the
major axis  of the central galaxy.  We restrict $\theta$  to the range
$0^{\circ}  \leq   \theta  <  90^{\circ}$,   where  $\theta=0^{\circ}$
($90^{\circ}$) implies that the satellite lies along the major (minor)
axis of the central galaxy. For a given set of centrals and satellites
we   then  count   the  total   number  of   central-satellite  pairs,
$N(\theta)$, for a number of  bins in $\theta$.  Next we construct 100
random samples  in which we  randomize the orientation of  all central
galaxies, and  we compute  $\langle N_R(\theta) \rangle$,  the average
number of central-satellite pairs  as function of $\theta$.  Note that
this ensures that  the random samples have exactly  the same selection
effects as the real sample, so that any significant difference between
$N(\theta)$  and  $\langle  N_R(\theta)  \rangle$ reflects  a  genuine
alignment  between the  orientation of  the central  galaxies  and the
distribution of satellite galaxies.

To quantify the  strength of any possible alignment  we follow Y06 and
define the distribution of normalized pair counts:
\begin{equation}
  P(\theta) = N(\theta) / \langle N_R(\theta) \rangle \,,
\end{equation}
Note  that  $P(\theta)=1$  in  the  absence of  any  alignment,  while
$P(\theta)>1$ at small $\theta$  implies a satellite distribution with
a preferred  alignment along the major  axis of their  centrals.  As a
measure   of   the   statistical   error   on   $P(\theta)$   we   use
$\sigma_R(\theta)    /    \langle    N_R(\theta)    \rangle$,    where
$\sigma_R(\theta)$ is the standard deviation of $N_R(\theta)$ obtained
from  the 100  random  samples.   We also  compute  the average  angle
$\langle  \theta  \rangle$.   Major  and  minor  axis  alignments  are
characterized by  $\langle \theta  \rangle < 45^{\circ}$  and $\langle
\theta  \rangle  >  45^{\circ}$, respectively\footnote{Note,  however,
  that  $\langle \theta  \rangle  = 45^{\circ}$  does not  necessarily
  imply  an   isotropic  distribution.   Therefore,   the  $P(\theta)$
  statistic  is  more  informative.}.   The significance  of  such  an
alignment can  be expressed in terms of  $\sigma_{\theta}$, defined as
the variance  in $\langle \theta  \rangle_R$ as obtained from  the 100
random samples.

\section{The Alignment in Dark Matter Haloes}
\label{sec:align}
\begin{figure}
\centerline{\psfig{figure=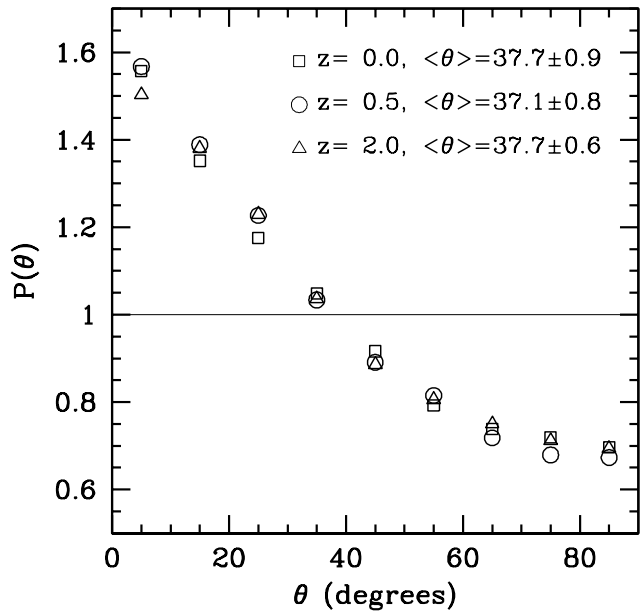,width=\hssize}}
\caption{The  normalized probability distribution $P(\theta)$ for
  central-satellite  pairs in  VIR haloes  at different  redshifts, as
  indicated. Clearly,  the alignment strength in the  SAM is virtually
  independent of redshift.}
\label{fig:time}
\end{figure}

In this section,  we first study the alignment  signal between central
and  satellite  galaxies  using  the distribution  of  galaxies  taken
directly  from the  simulation box  and projected  along  the $z$-axis
(arbitrary). Here we ignore all observational selection effects: we do
not model redshift-space distortions, nor do we consider a flux-limit.
In addition,  we consider  all central-satellite pairs  (with $M_{b_J}
\leq -16$)  that reside  in the  same dark matter  halo (with  $M \geq
10^{12} h^{-1} \Msun$). The analysis  of Y06, to which we will compare
our results, however,  is based on a flux-limited  redshift survey, in
which  centrals and  satellites are  grouped together  using  a galaxy
group finder. This results  in interlopers and incompleteness, which
are not  accounted for  here.  Rather, the  results presented  in this
section represent the true, uncontaminated alignment strengths present
in our (projected) simulation box.  In \S\ref{sec:mis} we will examine
the impact of observational  selection effects by using realistic mock
catalogues to which  we apply the same galaxy group  finder as used by
Y06.
\begin{figure*}
\centerline{\psfig{figure=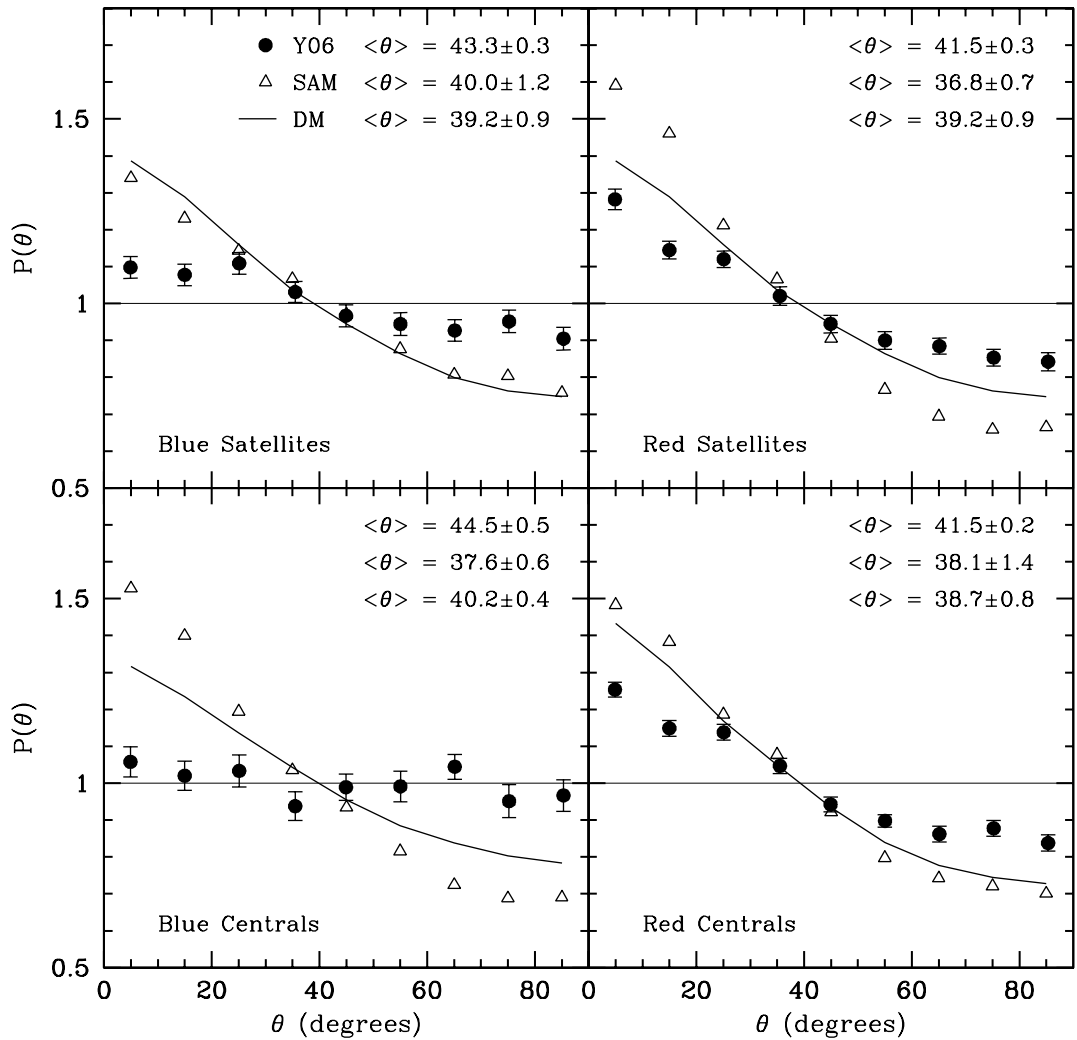,width=0.8\hdsize}}
\caption{The normalized   probability  distribution, $P(\theta)$, for
  various subsamples. The upper panels show the results for blue (left
  panel) and red (right panel) satellites, while the lower panels show
  the  results for  haloes with  blue  centrals (left  panel) and  red
  centrals (right panel).  In each  panel, the open triangles show the
  results for the satellite galaxies  in the SAM, the solid line shows
  the results for the dark matter  particles in the SAM, and the solid
  dots with errorbars show the observational results of Y06.}
\label{fig:color}
\end{figure*}

Fig.~\ref{fig:all} plots  the probability  distribution,  $P(\theta)$,
obtained  under  the  assumption that  the major  axis  of the central
galaxies is perfectly aligned   with the major  axis of  its projected
dark  matter  halo.   The open  triangles  correspond to   the results
obtained with FOF haloes, counting all satellites that are part of the
FOF halo.  As  one  can see, the probability  distribution $P(\theta)$
peaks  at small $\theta$, indicating   that the satellite galaxies are
distributed  preferentially along  the major axes   of their projected
dark matter haloes.  This is also evident from  the fact that $\langle
\theta  \rangle = 36.2^{\circ}  \pm  0.9^{\circ}$, which deviates from
the    case  of no   alignment   (i.e.,   $\langle  \theta  \rangle  =
45.0^{\circ}$) by almost $10\sigma$.   This alignment is simply due to
the non-spherical nature of  dark matter haloes and  to the  fact that
satellite galaxies  are a  reasonable    tracer of the  overall   mass
distribution  (e.g., Zentner \etal 2005;  Kang  \etal 2005b; Libeskind
\etal  2005;  AB06). The open squares  in  Fig.~\ref{fig:all} show the
alignment signal   obtained with the VIR   haloes,  only counting those
satellites with  $r < \Rvir$ Note  that the alignment signal for these
VIR  haloes is somewhat lower than   for the FOF  haloes.  This simply
owes  to  the fact  that  the VIR haloes   are confined to a spherical
radius. Give that  the observational results of  Y06 are also confined
to  a spherical    (group) radius,  and   that the  virial  masses are
physically better defined  than  the FOF masses,   in what follows  we
focus on the VIR haloes, unless specifically stated otherwise.
\begin{figure*}
\centerline{\psfig{figure=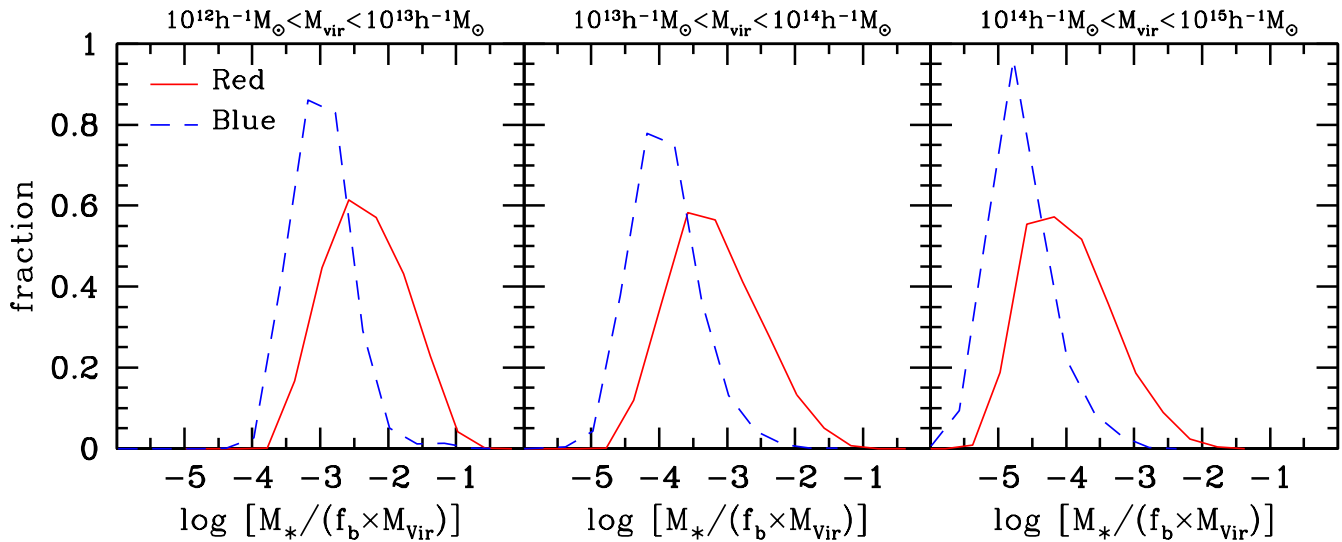,width=\hdsize}}
\caption{Distributions of the stellar masses of red (red, solid lines)
  and blue  satellites (blue, dashed  lines).  The stellar  masses are
  normalized  by the total  baryonic mass,  $f_b \times  M_{\rm vir}$,
  where $f_b=0.15$ is the  universal baryon fraction. The three panels
  correspond to different  bins in halo mass, as  indicated. Note that
  red  satellites  are,  on  average,  more massive  than  their  blue
  counterparts. As shown by  Libeskind \etal (2005), this implies that
  their subhalo masses, at the time of accretion, were more massive.}
\label{fig:stelmass}
\end{figure*}

The dashed and solid lines in  Fig.~\ref{fig:all} show the $P(\theta)$
distributions for the dark matter particles of the FOF and VIR haloes,
respectively.    Clearly  the dark  matter    particles also reveal an
alignment signal, though   it  is somewhat   weaker than  that of  the
satellite galaxies.   This suggests that satellite  galaxies are not a
perfect tracer of the dark matter distribution, but that in fact their
distribution is somewhat more flattened than that  of the halo itself.
This holds for both the  FOF and VIR haloes.   This is in  qualitative
agreement with  AB06, who also  noticed a  similar weak enhancement of
the alignment strength of satellites with respect  to the dark matter.
It is also in agreement with the simulation results of Libeskind \etal
(2005) and Zentner  (2005), who demonstrated that  (massive) subhaloes
(in Milky-Way  type haloes) tend to  be more strongly aligned with the
major axis of the host halo than the dark matter particles themselves.
This owes to the  preferred infall along  filaments, which tend  to be
preferentially aligned   with    the major axis of     the   halo.  In
\S\S~\ref{sec:galcol}   and~\ref{sec:halomass}   we   show  that  this
difference  between  the  $P(\theta)$ of  satellite galaxies  and dark
matter particles is  a function of halo mass  and satellite color, and
that it disappears for massive haloes.

Finally,  for comparison,  the solid  dots with  errorbars (reflecting
$\sigma_R(\theta)   /   \langle   N_R(\theta)   \rangle$)   show   the
observational results  obtained by Y06 from the  SDSS group catalogue,
using only those  groups with an inferred mass  $M \geq 10^{12} h^{-1}
\Msun$. With  $\langle \theta \rangle =  42.2^{\circ} \pm 0.2^{\circ}$
it is  clear that  the observed alignment  signal is much  weaker than
what is  obtained from our SAM  (see also AB06).   This indicates that
either (i) there are large observational selection effects that reduce
the strength of the alignment  signal, or (ii) that the orientation of
the central galaxies  is not perfectly aligned with  the major axis of
the projected host halo.  We  will test these two hypotheses in detail
in \S\ref{sec:mis}.

\subsection{Redshift Dependence}
\label{sec:zdep}
\begin{figure*}
\centerline{\psfig{figure=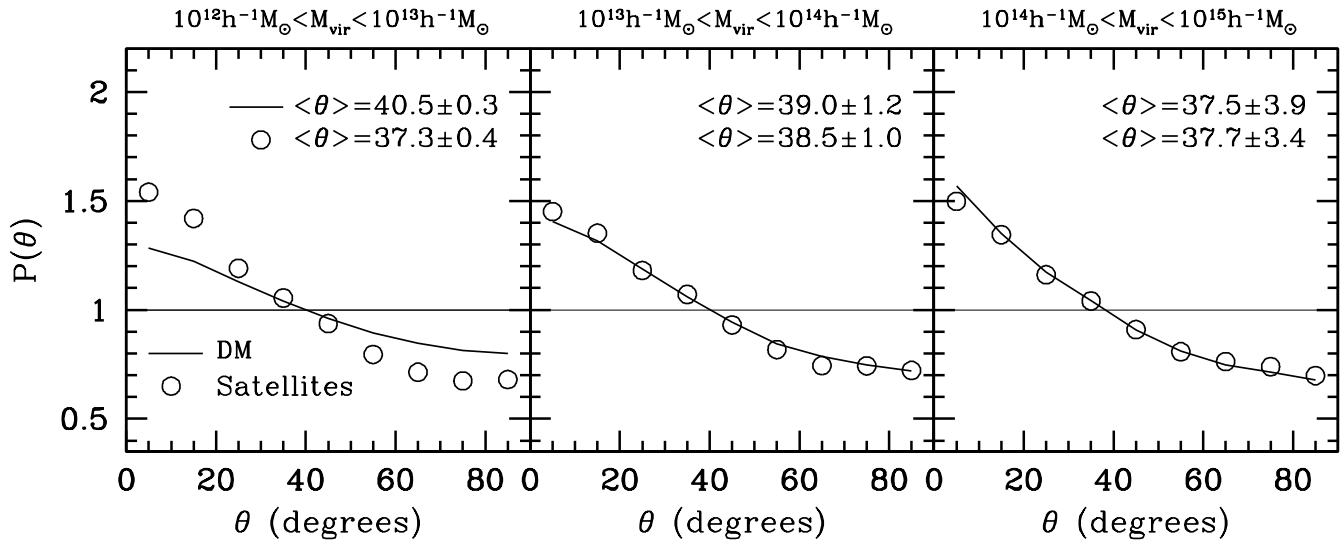,width=\hdsize}}
\caption{The same as  Fig.~\ref{fig:color}, but  for  galaxy pairs  in
  haloes  of  different masses.   The  open  circles  and solid  lines
  correspond  to the  satellite  galaxies and  dark matter  particles,
  respectively.  Note  the strong  difference between DM  and galaxies
  within halo with mass less than $10^{13}M_{\odot}$, galaxies in less
  massive haloes have stronger alignment that dark matter, but not the
  case for massive haloes.}
\label{fig:mass}
\end{figure*}

In  a recent study,  Donoso, O'Mill  \& Lambas  (2006) used  the Sloan
Digital  Sky Survey data  release 4  (Adelman-McCarthy \etal  2006) to
study the  alignment of  luminous red galaxies  at $z \sim  0.5$. They
found  that the major  axes of  these systems  are aligned  with their
surrounding galaxy  distributions, similar as  in the local  universe. 
Motivated by  these findings, we investigate how  the alignment signal
evolves  with  redshift  in  our   SAM.   The  results  are  shown  in
Fig.~\ref{fig:time}, where the triangles, circles and squares indicate
the model  predictions for VIR  haloes at $z=2.0$, $z=0.5$  and $z=0$,
respectively.   The  differences  between  the  alignment  signals  at
different redshifts are extremely small, with $\langle \theta \rangle$
that are all consistent with  each other at the $1\sigma$ level.  Note
that these  results are obtained  by selecting, at each  redshift, all
galaxies in  FOF haloes  with $M \geq  10^{12} h^{-1} \Msun$  and with
$M_{b_J}  \leq -16$.   In  a flux  limited  survey, however,  brighter
galaxies, which typically reside in more massive haloes, sample higher
redshifts.  If the alignment strength  depends on halo mass, as in the
data analyzed  by Y06 (see also \S\ref{sec:halomass}  below), then one
has to be  careful to properly separate redshift  dependence from halo
mass dependence.

\subsection{Dependence on Galaxy Color}
\label{sec:galcol}

We now examine how the alignment signal depends  on various galaxy and
halo  properties.   Fig.~\ref{fig:color} shows   the dependence of the
alignment   signal  on the colors   of   the satellite galaxies (upper
panels)  and the central galaxies  (lower panels).  The open triangles
show the  results  obtained from  our  SAM,   while the  observational
results of Y06 are shown as solid  dots.  Similar to the results shown
in Fig.~\ref{fig:all}, the SAM  yields much stronger alignment signals
than observed.  Again  we defer the  discussion of this difference and
its implications to \S\ref{sec:mis}.    Here  we simply focus on   the
color dependence.  First of all, the SAM  predicts that blue satellite
are less strongly aligned with the orientation of their central galaxy
than red  satellites,  which is  in   qualitative agreement with   the
observations.  The solid lines indicate  the $P(\theta)$ for the  dark
matter particles.  This  shows that blue  satellites  have  a $\theta$
distribution  that is virtually identical  to that of  the dark matter
particles,  while red  satellites reveal  an  alignment signal that is
clearly enhanced with respect to that of the dark matter.

In  order to gain  insight in  the origin  of this  enhanced alignment
signal of  red satellites, we have inspected  the galaxy distributions
in  the  SAM.   This  shows  that red  satellites  are  more  radially
concentrated than the blue  satellites, in agreement with observations
(e.g., Postman \& Geller 1984;  Girardi \etal 2003; Biviano \& Katgert
2004;  Thomas  \& Katgert  2006).   In  addition,  we find  that  blue
satellites  are  mostly  associated  with  {\it  halo}  galaxies  (the
satellites that  still have a  detectable subhalo around  them).  This
owes to  the fact  that most  of them have  been accreted  only fairly
recently.   Red  satellites,  on  the  other  hand,  have  their  star
formation largely  truncated and have been  accreted a long  time ago. 
Consequently, they  make up the  majority of orphans,  whose subhaloes
have  been  disrupted  by  tidal stripping.   The  crucial  difference
between  red  and blue  satellite  galaxies,  however, which  actually
explains their  different alignment strengths, is  their difference in
{\it  stellar}   mass.   As  shown   in  Fig.~\ref{fig:stelmass},  red
satellites are  on average significantly more massive  than their blue
counterparts.  As shown by Libeskind \etal (2005), the stellar mass of
a satellite galaxy is strongly correlated with the mass of its subhalo
{\it at  the time of accretion}.  Furthermore,  Libeskind \etal (2005)
and Wang  \etal (2005) have  clearly demonstrated that  subhaloes that
were  the most massive  at the  time of  accretion were  accreted more
preferentially  along  the  halo's  major axis.   Consequently,  their
distribution  is typically more  strongly flattened  than that  of the
other subhaloes  or than the  overall dark matter  distribution.  This
owes to  the fact that the  filamentary alignment of  the most massive
progenitors  is  largely  preserved  in  the final  halo,  and  nicely
explains  why the  red (more  massive) satellite  galaxies in  our SAM
reveal a stronger alignment signal  than either the dark matter or the
blue satellites.

The lower  panels of Fig.~\ref{fig:color}  show that the  SAM predicts
that  the satellite alignment  around blue  centrals is  comparable to
that around red centrals.  This differs from what it seen for the dark
matter  (solid lines) and  the results  of Y06  (solid dots),  both of
which show a significantly stronger alignment of satellites around red
centrals.  As we  show in  the next  section, the  fact  that $\langle
\theta \rangle$ for  the dark matter particles is  smaller around blue
centrals mainly owes  to the fact that blue  centrals mainly reside in
less  massive  haloes,  which  are more  spherical.   The  discrepancy
between the  alignment signals  in the SAM  and those obtained  by Y06
will be discussed in detail in~\S\ref{sec:mis}.

\subsection{Dependence on Halo Mass}
\label{sec:halomass}

Next  we  examine  how  the  alignment  scales  with  halo  mass.   In
Fig.~\ref{fig:mass}, we  plot $P(\theta)$ for  central-satellite pairs
in  different  halo mass  bins.   Before  looking  into the  alignment
signals of  the satellite  galaxies, let us  first focus on  the solid
lines, which  reflect $P(\theta)$ of the dark  matter particles.  Note
how  $\langle \theta  \rangle$ decreases  with increasing  halo  mass. 
This reflects  the well-known fact  that more massive haloes  are more
strongly  flattened (e.g., Warren  \etal 1992;  Bullock 2002;  Jing \&
Suto 2002;  Bailin \&  Steinmetz 2005; Kasun  \& Evrard  2005; Allgood
\etal 2006).  

For  haloes with  $M \gta  10^{13} h^{-1}  \Msun$ the  total satellite
population has a  $P(\theta)$ that is almost identical  to that of the
dark  matter particles,  indicating that  in massive  haloes satellite
galaxies are a fair tracer of the overall dark matter distribution (at
least  in the SAM  studied here).   In low  mass haloes,  however, the
satellite galaxies  reveal an  alignment signal that  is significantly
stronger  than   that  of  the  dark  matter   particles.   Thus,  the
differences between $P(\theta)$ of dark matter particles and satellite
galaxies  shown  in  Fig.~\ref{fig:all}  mainly owes  to  haloes  with
$M_{\rm  vir} \lta  10^{13}  h^{-1} \Msun$.   Note  that overall,  the
alignment signal  of the satellite  galaxies is almost  independent of
halo  mass:  the fact  that  more  massive  haloes are  more  strongly
flattened, is  roughly counter-balanced by  the fact that in  low mass
haloes, the  satellite galaxies are  more strongly flattened  than the
dark matter.

\subsection{Dependence on Luminosity}
\label{sec:lum}
\begin{figure}
\centerline{\psfig{figure=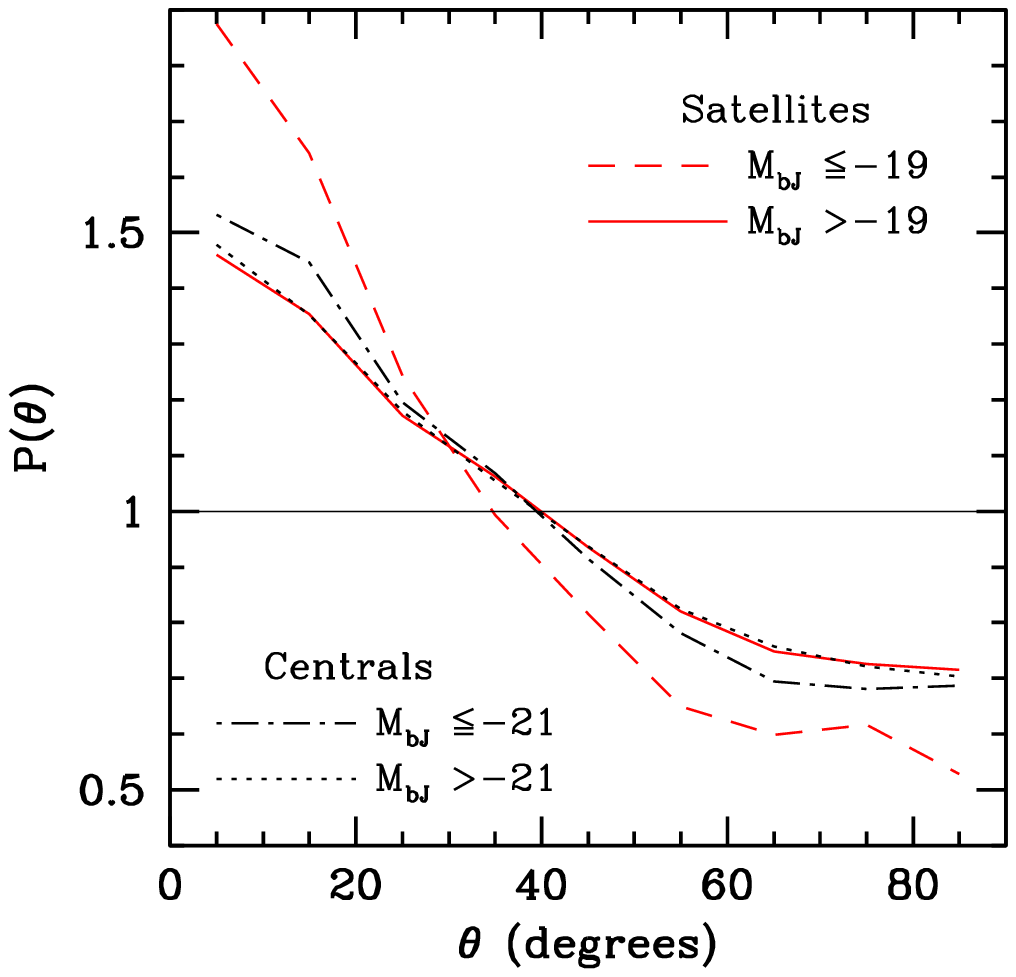,width=\hssize}}
\caption{The normalized probability distribution, $P(\theta)$, for VIR
  haloes  in  our SAM  for  two  absolute  magnitude limits  for  both
  satellites  and  centrals  respectively,  as  indicated.  When  only
  focusing on the brightest satellite galaxies, one obtains a stronger
  alignment  signal.   This  confirms  that more  massive  (and  hence
  brighter) satellite  galaxies are more  strongly flatten distributed
  than their lower mass counterparts.}
\label{fig:lum}
\end{figure}

Finally,  we  examine  how  the  alignment  strength  depends  on  the
luminosities   of  the   galaxies  (respectively   for   centrals  and
satellites). So far we always  considered all galaxies in the SAM with
$M_{b_J} \leq -16$.   Fig.~\ref{fig:lum} shows how $P(\theta)$ changes
if  we  change  this   magnitude  limit  for  satellites  and  centrals
respectively.   It is  found  that luminous  satellites have  stronger
alignment signal than faint  satellites, but the signals are identical
for  central galaxies regardless  of the  luminosity.  Since  there is
virtually no halo mass dependence, it  is obvious that why there is no
dependence  on the  luminosity  of the  central  galaxy.  Rather,  the
luminosity dependence  of $P(\theta)$ for satellites owes  to the fact
that more  massive (and hence  brighter) satellites reveal  a stronger
alignment signal; the same effect that is responsible for boosting the
alignment of  red satellites with  respect to that of  blue satellites
(cf.  \S\ref{sec:galcol}).   As we will see below,  this has important
implications  for the  alignment signal  inferred from  a flux-limited
sample of galaxies.

\section{The Alignment in Galaxy Groups}
\label{sec:mis}

As  illustrated   in  Figs.~\ref{fig:all}  and   \ref{fig:color},  the
alignment signals  found observationally by  Y06 are much  weaker than
those obtained from  our SAM. This, however, should  not come entirely
as  a surprise.  First  of all,  Yang \etal  studied the  alignment in
galaxy groups, not in dark  matter haloes.  Although the group finding
algorithm  used is  optimized to  group together  those  galaxies that
reside  in the  same dark  matter halo,  the groups  suffer  from both
interlopers (group  members that  do not actually  belong to  the same
halo), and incompleteness  (halo members that are missed  by the group
finder).   In addition, the  group catalogue  is constructed  from the
SDSS, which is a  flux-limited survey that suffers from incompleteness
itself   (i.e.,   fiber-collisions).    As   we  show   below,   these
observational  `selection'  effects  cause  a  blurring  of  the  true
alignment signal.   Furthermore, so far  we have oriented  the central
galaxies  along the  major axis  of  the {\it  projected} dark  matter
distribution, which maximizes the  alignment signal. However, since we
believe haloes to be triaxial, the projected major axis depends on the
viewing angles, making such an identification unphysical.

In this  section, we study  the impact of the  observational selection
effects,  and we  consider a  few more  realistic models  for  how the
central galaxy is oriented in its dark matter halo.

\subsection{Constructing mock catalogs}
\label{sec:mock}

In  order  to  examine  the  survey selection  effects,  we  start  by
constructing a mock SDSS DR2 galaxy sample.  A detailed description of
how the mock galaxy catalogue is  constructed can be found in Li \etal
(2006).  Here  we only give a  brief description.  Since  the SDSS DR2
survey extends  to redshifts $z \sim  0.3$, we need to  cover a volume
that extends to a radial distance  of about $900 h^{-1} \Mpc$. To that
extent we  create a $18 \times  18 \times 18$ periodic  replica of our
$100 h^{-1} \Mpc$ simulation box. Next we define an ecliptic ($\alpha,
\delta$)-coordinate frame  with respect to a  virtual observer located
in the  central box, and we  remove all galaxies outside  the SDSS DR2
survey regions.   For each galaxy  in the survey region,  we calculate
the redshift  using its comoving distance and  peculiar velocity along
the line of sight.  We  then calculate the $r$-band apparent magnitude
and  select  galaxies according  to  the position-dependent  magnitude
limit  in  the  SDSS   DR2.   We  also  mimic  the  position-dependent
completeness using the completeness masks provided as part of the SDSS
DR2.  Finally, we select those galaxies with $0.01 < z < 0.2$

Note that  in the  spirit of constructing  the mock survey  above, for
each mock  galaxy, we know exactly  the identity of its  halo from our
simulation  described in  detail  in  Section 2.1,  and  we also  know
whether it  is a  central or satellite  of that  halo. Now we  use the
resulting mock survey to  study the central-satellite alignments using
two different methods.   In the first method, we make  use of the fact
that for each galaxy in the mock survey, we know its host halo, and we
know whether  the galaxy  is a  central galaxy, a  halo galaxy,  or an
orphan galaxy from our N-body  simulation. It is then easy to allocate
galaxies into different haloes.  In case a halo without central galaxy
selected by the SDSS selection effect, we omit all the galaxies in the
halo. Overall  the fraction  of such haloes  is very small.   For each
halo with a  central galaxy we project the  dark matter halo particles
(taken from  N-body simulation) onto the  `sky' and we  use the method
mentioned  in \S\ref{sec:orient}  to  determine its  major axis  (once
again in projection). As before,  we assume that the central galaxy is
aligned  along  this major  axis,  and  we  determine $P(\theta)$  and
$\langle  \theta \rangle$  using all  satellite galaxies  in  the mock
survey  that are  located  within the  same  VIR halo  as the  central
galaxy.  The halo  mass and radius ($\Rvir$) of  the central galaxy is
also taken directly from the  simulation.  In what follows we refer to
the results  obtained in this way  as the {\it Mock  Halo} results.  A
comparison of these results  with those discussed in \S\ref{sec:align}
above reveals the impact of  a flux-limit, of peculiar velocities, and
of the incompleteness of the SDSS on the alignment signal.

In the second method we aim  for a more meaningful comparison with the
results of Y06  by applying their halo-based group  finder to our mock
survey to construct a mock  galaxy group catalog. Following Yang \etal
(2005a),  we compute  a measure  for the  total group  luminosities by
summing the luminosities of its member galaxies, and by correcting for
missing members using a calibration based on relatively nearby groups.
We then obtain group masses  by matching the group luminosity function
to the (theoretical) halo  mass function assuming a monotonic relation
between group  luminosity and  halo mass (see  Yang \etal  2005a,b and
Weinmann  \etal  2006 for  details  and  for  tests demonstrating  the
reliability of  the assigned group masses).   Note here we  do not use
the  identify of  the  galaxy  (central or  satellite)  in the  N-body
simulation. Following Y06, we define the brightest group member as the
`central' galaxy, and all other group members as `satellites'. We thus
obtain   a   sample   of   $12418$  central   galaxies   and   $37011$
central-satellite  pairs,  comparable  to  Y06 ($16013$  centrals  and
$39086$  central-satellite pairs).  We  align the  major axis  of each
`central'  group galaxy  with the  major  axis of  its projected  dark
matter  halo (VIR),  and we  then determine  $P(\theta)$  and $\langle
\theta \rangle$ using all other group members as satellites.  We refer
to the results  obtained in this way as the  {\it Mock Group} results.
Note  that  this  is exactly  the  same  method  as  used by  Y06.   A
comparison with  the results  presented in \S\ref{sec:align}  and with
the {\it  Mock Halo}  results thus  allows us to  study the  impact of
interlopers (group members  that are not located in  the same VIR halo
as  the  central group  galaxy),  incompleteness,  and  errors in  the
assigned group mass.

\subsection{The Impact of Selection Effects}
\label{sec:select}
\begin{figure}
\centerline{\psfig{figure=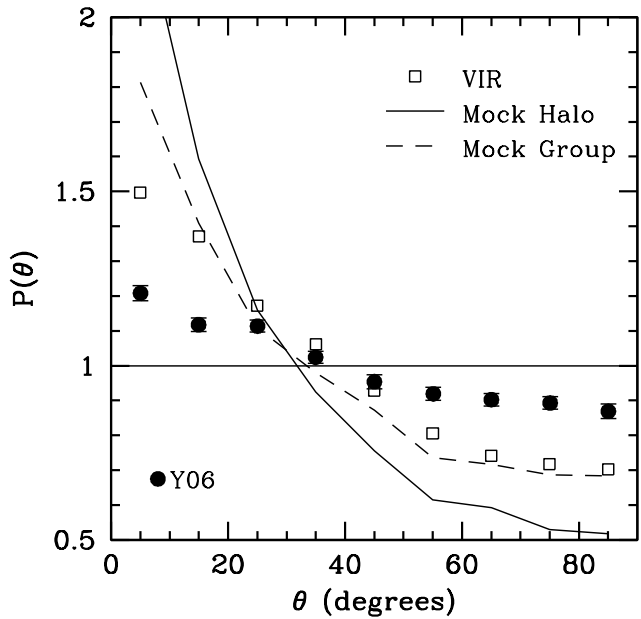,width=\hssize}}
\caption{The alignment signal obtained from our {\it Mock Halo} (solid 
  line) and {\it Mock Group} (dashed line) samples, where as before we
  have assumed that  the central galaxy is perfectly  aligned with the
  major axis  of its  projected dark matter  halo. For  comparison, we
  also  show  the  results for  the  VIR  haloes  (open squares,  cf.  
  Fig.~\ref{fig:all}) and the observational results of Y06 (solid dots
  with errorbars).}
\label{fig:mis1}
\end{figure}

\begin{figure}
\centerline{\psfig{figure=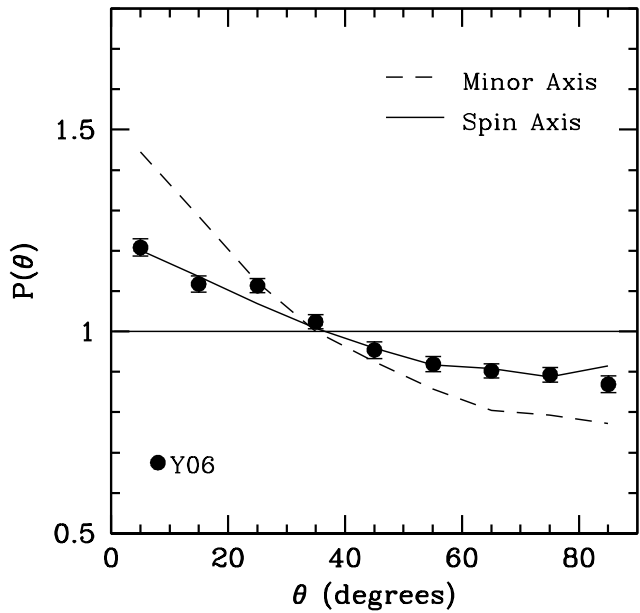,width=\hssize}}
\caption{The alignment signal obtained from our {\it Mock Group}
  sample for  two different alignment models: The  dashed (solid) line
  corresponds to a model in which we assume that the minor axis of the
  central galaxy  is perfectly aligned  the minor axis (spin  axis) of
  its  dark   matter  halo.   Whereas  the  {\it   Minor  Axis}  model
  overpredicts the  alignment signal obtained by Y06  (solid dots with
  errorbars),  the {\it  Spin Axis}  model accurately  fits  the data,
  suggesting that  the orientation of central galaxies  is governed by
  the angular momentum vector of their dark matter halo.}
\label{fig:mis2}
\end{figure}

The  solid and dashed  lines in  Fig.~\ref{fig:mis1} show  the results
obtained  from   the  {\it  Mock   Haloes}  and  {\it   Mock  Groups},
respectively.  In  both cases we only  use haloes (or  groups) with $M
\geq 10^{12} h^{-1} \Msun$.  For  comparison, we also show the results
obtained in  \S\ref{sec:align} for the  VIR haloes (open squares,  cf. 
Fig~\ref{fig:all}) and  the observational  results of Y06  (solid dots
with errorbars).   The alignment signal  for the {\it Mock  Haloes} is
actually  stronger than  what  we obtained  using  the simulation  box
directly.   This is  due  to the  fact  that the  mock catalogues  are
apparent   magnitude   limited.    Consequently,   brighter   galaxies
contribute a relatively larger fraction to the total alignment signal.
As we have shown in \S\ref{sec:lum}, brighter (more massive) satellite
galaxies reveal  a stronger alignment signal than  their fainter (less
massive) counterparts (cf. Fig.~\ref{fig:lum}).

Note  also that the  alignment signal  from the  {\it Mock  Groups} is
significantly weaker than for the {\it Mock Haloes}.  This mainly owes
to fact that $\sim 20$ percent  of the satellite galaxies in the group
catalogue are  interlopers. Since interlopers are  not associated with
the halo  of the central,  they tend to  dilute the alignment  signal. 
Since our mock group catalogue  and the group catalogue constructed by
Y06 should be impacted by interlopers in roughly the same fashion, the
data-model  comparison is still  valid. Therefore,  the fact  that the
alignment signal in  the {\it Mock Group} catalogue,  which mimics all
observational  selection effects,  is  still much  stronger than  that
obtained  by Y06,  implies  that central  galaxies  are not  perfectly
aligned with the  orientation of their projected dark  matter halo. In
the following section we  use more realistic assumptions regarding the
alignment between central galaxies and their dark matter haloes.

\subsection{Different Alignment Models}
\label{sec:diffmod}
\begin{figure*}
\centerline{\psfig{figure=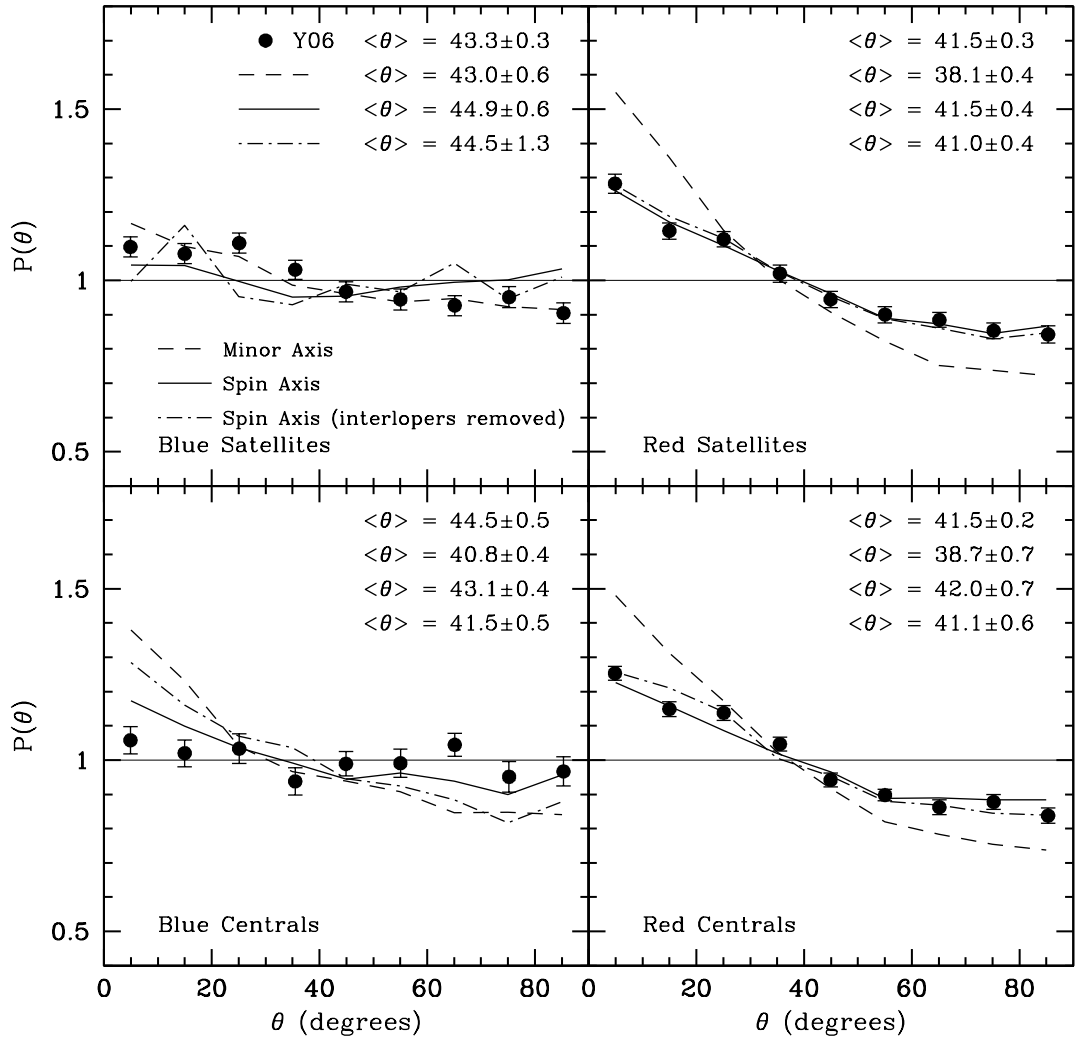,width=0.8\hdsize}}
\caption{Same as Fig.~\ref{fig:color}, but for the galaxies in our
  {\it Mock Group} sample.  Results  are shown for both the {\it Minor
    Axis} model  (dashed lines) and  the {\it Spin Axis}  model (solid
  lines).  Note that the latter  provides a remarkably good fit to the
  data of  Y06 (solid dots  with errorbars).  For comparison,  we also
  show the  results for  the {\it  Spin Axis} model  in which  we have
  manually removed the interlopers from the {\it Mock Group} catalogue
  (dot-dashed lines). A comparison with the solid lines highlights the
  impact of interlopers. See text for a detailed discussion.}
\label{fig:model}
\end{figure*}

In what  follows we assume  that the central  galaxy is a  pure oblate
spheroid  (of which  a  thin disk  is  the extreme  case,  as used  by
AB06).  In addition,  we  assume that  its  minor axis  is either  (i)
perfectly  aligned  with  the  minor  axis of  its  dark  matter  halo
(hereafter {\it Minor Axis} model), or (ii) perfectly aligned with the
angular momentum vector  of its dark matter halo  (hereafter {\it Spin
Axis} model).In both case we use  the VIR haloes, and we determine the
halo's 3D inertia  tensor and angular momentum vector  as described in
\S\ref{sec:orient}.  Following  AB06, we  project oblate into  the sky
and obtain the major axis of the image. the As long as the halo itself
is  a spheroid  (oblate or  prolate), the  {\it Minor  Axis}  model is
identical to the case discussed above,  in which the major axis of the
central galaxy is  perfectly aligned with that of  its {\it projected}
dark  matter halo.   If, however,  the halo  is triaxial,  this  is no
longer true in general.  The {\it Spin Axis} model is motivated by the
standard model  for disk formation, in  which the disk forms  out of a
cooling flow that conserves  its specific angular momentum (e.g., Fall
\& Efstathiou 1980; Mo, Mao \&  White 1998; van den Bosch 2001; Dutton
\etal 2007).  If  the baryons and dark matter start  out with the same
specific  angular momentum, which  is a  standard assumption  in these
models (but see van den Bosch \etal 2002; Chen, Jing \& Yoshikaw 2003;
Sharma \&  Steinmetz 2005), then  the angular momentum vectors  of the
disk and halo should be aligned.

Fig.~\ref{fig:mis2}  shows  $P(\theta)$ obtained  from  the {\it  Mock
  Group} catalogue,  for both  the {\it Minor  Axis} (solid  line) and
{\it Spin  Axis} (dashed  line) models.  First  of all, note  that the
{\it Minor Axis} model yields an alignment strength that is much lower
than  that for  the {\it  Mock Groups}  in  Fig.~\ref{fig:mis1}.  This
shows  that overall  dark matter  haloes are  triaxial.   However, the
alignment signal  is still  significantly larger than  observed (solid
dots with errorbars),  indicating that the {\it Minor  Axis} model can
not represent reality.  The {\it  Spin Axis} model, on the other hand,
can accurately  reproduce the  satellite alignment signal  obtained by
Y06.   Therefore, the data  is consistent  with a  model in  which the
central galaxy is  oblate and perfectly aligned with  the spin axis of
its  dark matter  halo.  
\begin{figure*}
\centerline{\psfig{figure=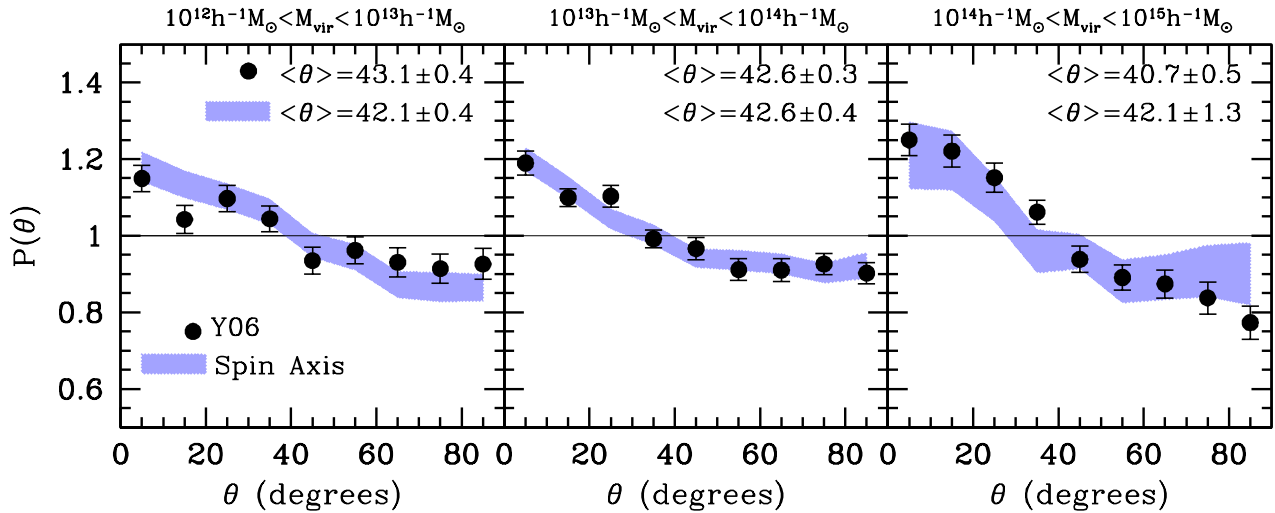,width=0.95\hdsize}}
\caption{The alignment signal $P(\theta)$ obtained from our 
  {\it Mock  Group} catalogue  for the {\it  Spin Axis}  model (shaded
  bands) for three bins in  group mass (as indicated), compared to the
  results of Y06.}
\label{fig:hmass}
\end{figure*}

The  fact that the  {\it Spin  Axis} model  yields a  weaker satellite
alignment  signal than  the {\it  Minor Axis}  model implies  that, in
general, the  spin axis and minor  axis of dark matter  haloes are not
perfectly  aligned.  This  is consistent  with  numerical simulations,
which show  that although the angular  momentum axis of  a dark matter
halo  tends to  be preferentially  aligned  with the  minor axis,  the
alignment is far from perfect.   For the VIR haloes in our simulation,
we find  an average angle (in  3D) between the minor  axis and angular
momentum  axis of  $43.2^{\circ}$, in  good agreement  with  Bailin \&
Steinmetz (2005) and Bett \etal (2006).

To  test the  {\it Minor  Axis}  and {\it  Spin Axis}  models in  more
detail, Fig.~\ref{fig:model} shows  $P(\theta)$ obtained from the {\it
  Mock Group} catalogue for blue and red satellites (upper panels) and
for blue and  red centrals (lower panels). The  {\it Minor Axis} model
only provides a  good fit to the Y06 data for  the blue satellites. In
all other  cases it significantly overpredicts  the observed alignment
signal.  The {\it Spin Axis} model, on the other hand, matches the Y06
results remarkably well  in all cases, providing strong  support for a
picture in  which the orientation  of central galaxies is  governed by
the angular momentum vector of their dark matter halo.

In order to study the impact of interlopers, the dot-dashed lines show
the results  for the {\it Spin  Axis} model in which  we have manually
removed  the interlopers  from the  group catalogue.   The differences
with the  solid lines therefore  highlight the impact of  interlopers. 
Overall the  differences are small;  only the alignment signal  of the
blue centrals seems to have been significantly diluted by interlopers.
This  owes to the  fact that  blue centrals  have a  larger interloper
fraction ($\sim  35$ percent) than  red centrals ($\sim 15$  percent). 
This in turn owes to the fact that group masses are estimated from the
group luminosities.  At  the low mass end, where  the group luminosity
is dominated  by the luminosity  of the central galaxy,  blue galaxies
typically get  an assigned mass which  is somewhat too  high (see More
\etal 2006, in preparation).  Consequently, the assigned virial radius
is  somewhat  too  large,  which  results  in  a  larger  fraction  of
interlopers.  Since the $P(\theta)$ for blue centrals and red centrals
are    similar   when    the    interlopers   are    removed   (cf.    
Fig.~\ref{fig:color}), we conclude that  the finding by Y06, that blue
centrals  are more  strongly aligned  with their  satellites  than red
centrals,  is most likely  an artefact  of the  method used  to assign
masses to the groups.

The  difference   between  the  alignment  signal  of   blue  and  red
satellites,  on the  other hand,  seems to  be a  genuine  effect, not
significantly   distorted  by  interlopers.    Indeed,  as   shown  in
\S\ref{sec:galcol}, this difference is also evident when analyzing the
simulation box  directly, and is due  to the fact  that red satellites
are associated  with subhaloes that were  more massive at  the time of
accretion. We  therefore conclude  that the enhanced  alignment signal
for red satellites  has a natural explanation within  the framework of
hierarchical structure formation.

Finally, Fig.~\ref{fig:hmass}  compares the alignment  signal obtained
from our{\it Mock  Group} catalogue with the {\it  Spin Axis} model to
the data of Y06 for three  different bins in group mass.  The width of
the  shaded  band  reflects  $\sigma_R(\theta) /  \langle  N_R(\theta)
\rangle$, and  is shown  to highlight the  random scatter in  our {\it
  Mock   Group}  catalogue.    Consistent  with   what  we   found  in
\S\ref{sec:halomass}, the  alignment signal from our  {\it Mock Group}
catalogue  reveals no  significant mass  dependence. Although  the Y06
data seems to hint towards a decrease of $\langle \theta \rangle$ with
increasing  group   mass,  the   overall  agreement  with   our  model
predictions  is very  satisfactory,  providing further  support for  a
picture in which  the central galaxy is aligned with  the spin axis of
its dark matter halo.

\section{Conclusions}
\label{sec:concl}

We  have used a  high-resolution $N$-body  simulation combined  with a
semi-analytical  model for  galaxy  formation to  study the  alignment
between the  orientation of central  galaxies and the  distribution of
their satellite  galaxies.  Since dark matter haloes,  in general, are
flattened, and satellite galaxies are  a reasonably fair tracer of the
dark  matter  mass distribution,  satellite  galaxies  will reveal  an
alignment signal as  long as the orientation of  the central galaxy is
correlated  with that  of its  dark matter  halo.  In  particular, the
major  axis alignment  detected  by several  recent studies  (Brainerd
2005; Yang \etal 2006; Azzaro \etal 2006) requires that the major axis
of a central galaxy is somehow aligned with the major axis of its dark
matter halo.  Using our simulation we have constructed a mock SDSS, to
which we  applied the  same halo-based  group finder as  used by  Y06. 
Using  exactly  the same  analysis  as Y06,  we  find  that a  perfect
alignment of  the minor  axes of central  galaxy and dark  matter halo
yields  a  central-satellite alignment  signal  that is  significantly
stronger than observed.  However, if  we assume that the minor axis of
the central galaxy is perfectly aligned with the spin axis of its dark
matter halo,  which has  an average misalignment  with the  halo minor
axis  of $43.2^{\circ}$, we  can accurately  reproduce the  results of
Y06.

AB06 have  also obtained similar conclusions as  presented here. First
they also  found that if the  major axis of the  central galaxy aligns
well with the major axis of the projected dark matter halo, the signal
from the  model is  stronger than observed.   On the contrary,  if the
spin axis of  the central galaxy aligns with  the net angular momentum
of the dark  matter halo, the signal is  decreased.  AB06 claimed that
the signal  is decreased  too much, and  is lower than  observation of
Brainerd (2005).   Here we  find that our  Spin Axis model  match well
with  the  observations  of  Y06.   It  is  deserved  to  clarify  the
disagreement.   First,  AB06 compared  their  model  results with  the
observation  results  of  Brainerd  2005,  who  measured  the  average
alignment  angle as $40.5^{\circ}$,  but Y06  obtained $42.2^{\circ}$.
The difference in the two  observational results owes to the fact that
AB06 and  Y06 measured  the alignment signal  in different  halo mass.
AB06 measured the  signal mostly from isolated host  galaxies, but Y06
measured the  signal from  groups and clusters.   Second, in  fact our
result from Spin Axis model is  almost identical to that of AB06, here
we get  average alignment  angle of $42.4^{\circ}$,  and $42^{\circ}$
from AB06.  Third,  there is no statistical errorbars  from AB06, then
it is  hard to  conclude whether there  is a  significant disagreement
between  their  results  and  the observational  results  of  Brainerd
(2005).
    
The  analysis of  SDSS groups  by  Y06 revealed  a stronger  alignment
signal for red satellites than  for blue satellites (see also Sales \&
Lambas 2004 and Azzaro \etal 2006). This is remarkably well reproduced
by our SAM, and is related to the fact that red satellites have larger
stellar masses  (when normalized  by the mass  of the host  halo) than
their blue  counterparts, and are associated with  subhaloes that were
more massive at the time of  accretion.  As shown by Wang \etal (2005)
and  Libeskind \etal (2005),  the distribution  of those  subhaloes is
more strongly flattened than that of less massive subhaloes or that of
the dark matter particles.  This owes to the fact that the filamentary
alignment of the most massive  progenitors is largely preserved in the
final halo.  When assuming that  the minor axes of central galaxies is
perfectly  aligned  with their  halo  spin  axes,  our SAM  accurately
matches the data  of Y06, and we therefore  conclude that the enhanced
alignment signal  of red satellites  has a natural explanation  in the
framework of hierarchical structure formation.

The analysis of Y06 also  revealed a stronger alignment signal for red
centrals than for  blue centrals.  Although our {\it  Spin Axis} model
can  accurately  reproduce  this  trend  when  using  the  mock  group
catalogue, we find no enhanced  alignment signal for red centrals when
analyzing  the dark  matter haloes  in  the simulation  box directly.  
Detailed  tests  show that  blue  centrals  have  a larger  interloper
fraction, which causes  an enhanced dilution of the  alignment signal. 
We therefore conclude  that the difference in the  alignment signal of
red  and blue  centrals detected  by Y06  is most  likely  an artefact
caused by interlopers in the group catalogue used.

Finally, the alignment signal in the semi-analytical model is found to
only depend very weakly on halo  mass, in good agreement with the data
of Y06.  This lack of  a clear mass dependence is somewhat surprising,
since  it is well  known that  more massive  haloes are  more strongly
flattened.     Consequently,    one    would   expect    a    stronger
central-satellite   alignment  in   more  massive   haloes.   However,
satellites  in  low mass  haloes  are  relatively  more massive  (with
respect  to their  host halo).   Since  the alignment  signal of  more
massive   satellites   is    stronger,   this   counter-balances   the
mass-dependence  of  the  halo  shapes: Massive  haloes  are  strongly
flattened, and  their satellites accurately trace the  mass.  Low mass
haloes,  however, are  less  strongly flattened,  but their  satellite
distribution is more strongly flattened than the dark matter. 

Our  main  conclusion  is  that  the observed  alignment  between  the
orientation  of  central  galaxies   and  the  distribution  of  their
satellite galaxies is in good agreement with the standard hierarchical
structure  formation model,  as  long  as the  minor  axes of  central
galaxies  are misaligned  with the  minor  axes of  their dark  matter
haloes by on average $\sim 40^{\circ}$. Interestingly, this is exactly
the typical  misalignment angle  between a halo's  minor axis  and its
angular momentum vector, which is suggestive of a picture in which the
orientation of central galaxies is  governed by the spin axes of their
dark matter haloes.

At  first sight,  it  may seem  weird  that the  orientation of  (red)
elliptical galaxies would  be governed by the spin  axes of their dark
matter haloes.  However, except for the most massive ellipticals, more
early-type  galaxies are  oblate  rotators, which  implies that  their
flattening owes to  their angular momentum (e.g., Davies  \etal 1983). 
If ellipticals form in major  mergers, which is the standard paradigm,
this angular momentum originates  from the orbital angular momentum of
the  merging progenitors.  Interestingly,  several studies  have shown
that the angular  momentum of a dark matter  halo also originates from
the orbital  angular momenta of its (most  massive) progenitors (e.g.,
Vitvitska \etal  2002; Maller, Dekel \&  Somerville 2002).  Therefore,
we argue that  an alignment between the minor  axes of ellipticals and
the spin  axes of  their dark  matter haloes may  well have  a natural
origin  in hierarchical models  of structure  formation.  This  can be
tested straightforwardly with high-resolution numerical simulations.

For (blue) disk  galaxies, a tight alignment between  the spin axes of
halo and  disk seems a  natural outcome if  (i) the spin axes  of dark
matter and  baryons are initially  aligned and (ii)  cooling preserves
specific angular momentum.   However, hydrodynamic simulations suggest
that the angular  momentum vectors of the baryons  and dark matter are
not  perfectly aligned  (van den  Bosch \etal  2002; Chen  \etal 2003;
Sharma \& Steinmetz 2005).   Furthermore, detailed simulations of disk
formation show that the actual formation of the disk strongly modifies
the  shape  of  the  inner   dark  matter  halo  ($\lta  0.1  \Rvir$),
reorienting  it so  that the  minor  axis of  the inner  halo is  well
aligned with the disk axis (Kazantzidis \etal 2004; Bailin \etal 2005;
see also  Binney, Jiang \& Dutta  1998).  Such a  reorientation of the
inner  halo may  have an  impact on  the  central-satellite alignment,
which has not been accounted for  in our model.  It remains to be seen
to  what  extent  a  self-consistent  treatment of  the  formation  of
(central) galaxies, that accounts for  the back reaction of cooling on
the shape and orientation of the halo, impacts on the alignment signal
studied  here. In  particular, since  the radial  distribution  of red
satellites  is   more  centrally   concentrated  than  that   of  blue
satellites, the  impact of this effect  may well be  different for red
and blue satellites.

As a  final remark, we caution  that, although our  data suggests that
the orientation  of central galaxies is  governed by the  spin axes of
their  dark matter  haloes,  and there  are  theoretical arguments  to
support such a picture, any  alternative model in which the minor axes
of central galaxy and dark matter halo have an average misalignment of
$\sim 40^{\circ}$, will match the  data equally well. In that respect,
the origin  of the alignment between centrals  and satellites requires
further study.


\section*{Acknowledgements}

We thank  the anonymous referee for useful  comments.  XK acknowledges
support  from  the  Royal  Society  China  Fellowship  scheme.  XY  is
supported  by the  {\it One  Hundred Talents}  project of  the Chinese
Academy  of Sciences and  grants from  NSFC (Nos.10533030,  10673023). 
YPJ  is supported  by the  grants from  NSFC  (Nos.10125314, 10373012,
10533030) and  from  Shanghai Key  Projects  in  Basic  research (No.  
04JC14079 and  05XD14019).  HJM  and SM thank  the Chinese  Academy of
Sciences and Chinese Natural Science Foundation for travel support.



\label{lastpage}

\end{document}